\documentclass[aps,pra]{revtex4}
\usepackage{graphicx}

\def\bra#1{\langle #1 |}
\def\ket#1{| #1\rangle}

\def\ul{\underline}

\def\R{\hbox{\rm I \kern-5pt R}}

\def\Tr{{\rm{Tr}}}
\begin{document}
\title{Nonlinearity without Superluminality}
\author{Adrian Kent}
\email{a.p.a.kent@damtp.cam.ac.uk}
\affiliation{Hewlett-Packard Laboratories, Filton Road, Stoke Gifford, 
Bristol, BS34 8QZ, U.K.}
\altaffiliation[Present and permanent address: \qquad]{
Centre for Quantum Computation, Department of Applied Mathematics
and Theoretical Physics, University of Cambridge,
Wilberforce Road, Cambridge, CB3 OWA, U.K.}
\date{April 2002 (revised February 2005)}

\begin{abstract}
Quantum theory is compatible with special relativity. 
In particular, though measurements on entangled systems 
are correlated in a way that cannot be reproduced by local
hidden variables, they cannot be used for superluminal 
signalling.  As Czachor, Gisin, and Polchinski pointed out, this is 
not generally true of general nonlinear modifications of the Schrodinger
equation.  Excluding superluminal signalling has thus been taken  
to rule out most nonlinear versions of quantum theory. 
The no superluminal signalling constraint has also been 
used for alternative derivations of the optimal fidelities
attainable for imperfect quantum cloning and other operations. 

These results apply to theories satisfying the rule that their predictions 
for widely separated and slowly moving entangled systems can 
be approximated by non-relativistic 
equations of motion with respect to a preferred time coordinate.  
This paper describes a natural way in which this rule might
fail to hold.  In particular, it is shown that quantum readout devices which 
display the values of localised pure states need not 
allow superluminal signalling, provided that the devices display the values
of the states of entangled subsystems as defined in a non-standard, 
although natural, way.  
It follows that any locally defined nonlinear evolution of pure states can be
made consistent with Minkowski causality. 
\end{abstract}
\maketitle

\section{Motivations} 

There are at least three good reasons to look for alternatives to
quantum theory: the measurement problem, the difficulty in reconciling
quantum theory with general relativity, and the desirability of
finding new classes of theories against which certain quantum
principles, such as linearity, can be tested.  Yet it has proved
rather difficult to find alternatives to quantum theory which respect
the relativity principle and do not allow some form of superluminal
signalling.  For this and other reasons, the subtle relationship
between quantum theory and special relativity is a source of
continuing fascination.

Special relativity is not necessarily sacrosanct, of course, and
moreover superluminal signalling need not be inconsistent
with the relativity principle \cite{aksignal}.  
But the motivations just given 
suggest that alternatives to quantum theory
which respect the relativity principle and do not allow superluminal
signalling may be especially interesting and valuable \cite{hartle}.  If the aim is
to unify quantum theory and general relativity, abandoning the
relativity principle or Minkowski causality seems an unpromising
start.  Also, one would prefer to test principles such as linearity by
varying as little else as possible.  For instance, if a test confirms
a theory which respects linearity and relativity against a theory
which respects neither, it is not so clear whether to interpret this
as a confirmation of linearity or of relativity.  And then, the very
fact that respecting relativity and Minkowski causality seems to be
difficult could be a hint that it is necessary.  Constraints which are
difficult (but not impossible) to satisfy are particularly
interesting, since it would be nice to believe that the fundamental
theory of nature is defined by a few compelling principles, rather
than chosen arbitrarily from a large class of equally plausible
possibilities.

All these points suggest reconsidering the relation
between quantum theory and relativity. 

\section{Zweisteine's state readout machine} 

Your colleague Zweisteine has long been a zealous admirer of special
and general relativity but robustly sceptical about quantum theory.
He reserves a special venom for the treatment of measurement within
quantum theory.  Naturally, the imprecision of the notion of
measurement has not escaped his attention, and he believes that
quantum theory needs to be augmented by a precise theory of 
state reduction.  But he maintains also a
less widely held view.  He feels it is inconceivable that nature can
have created objects so subtly intricate as quantum
states, in such a form that we can access them only by the
brutally destructive process encapsulated in the projection postulate.
Positive operator valued measurements make him no happier: he
sees them merely as projections applied to a larger Hilbert space, 
bringing essentially the same unsatisfactory tradeoff between 
limited information gain and significant disturbance.

It must, he believes, be possible to access the information encoded in
a state more directly and less destructively.  Accordingly, he has for
some years been working on a quantum state readout machine.  This is
supposed to accept a qubit --- Mark I will be 
restricted to two dimensional systems --- which it returns unaltered
after printing out a high precision description. 
  
Quite some time ago, you drew his attention to the
no-cloning theorem \cite{wz} and related work \cite{day}.
He replied that these results
illuminate very elegantly the limitations of quantum theory, and more
generally the poverty of a universe limited to unitary or linear
evolution laws.  Fortunately, he added with an admonitory wag
of the finger, we know from general relativity that nature is 
essentially non-linear.

More recently, after a particularly fraught departmental
meeting, you were tactless enough to mention 
various papers that discuss the relation of quantum nonlinearity to 
superluminal signalling \cite{czachor,gisinrel,polchinski,sbg}
and even query whether a natural construction of nonlinear theories is
possible \cite{mielnik}.  
These cumulatively cast him into a state of
great gloom, from which even the visit of an eminent 
Everettian, with whom he would normally have delighted in
fencing, failed to rouse him.  

Yet today, the spring is again in his step, a gleam of triumph in his
eye.  He has seen a way around the no-superluminal signalling
constraint, he announces, and his state readout machine is complete.  
What can you do but indulge him?  You prepare a qubit in state $\ket{\psi} = a
\ket{0} + b \ket{1}$ in your lab, $a$ being positive real and $b$
complex, each specified to several decimal places.  You bring it
across, feed it into the machine.  The printout reads $ a \ket{0} + b
\ket{1}$.  You test the returned qubit, measuring $P_{\psi}$, and
get the answer $1$.  A lucky guess, perhaps.  After several similar
experiments, though, another explanation seems required.  

Whatever trickery is afoot, you know how to expose it.  
Your old colleague Isabelle, now based on Callisto, is happy to 
assist.  This evening, she prepares a pair of particles in a singlet
state, 
$$ (1 / \sqrt{2} ) ( \ket{0} \ket{1} - \ket{1} \ket{0} ) $$
and sends you the second particle.  
At noon tomorrow, universal time, she will carry
out a projective measurement, in a basis of her choice, on the
first particle.  
If she then reported the basis and result immediately by radio, the
signal would reach you at 1pm.  Guided by some faint premonition,
though, you ask her to delay sending the signal for half an hour.

The next morning, you feed the entangled qubit into Zweisteine's 
machine.  It whirrs, while you watch in amusement, and then 
prints out something surprising: 
$$ 1/2 \ket{0} \bra{0} + 1/2 \ket{1} \bra{1} \, . $$ 
Taking the returned qubit, you wait till $12.01$, for the 
crucial test, and resubmit the qubit.  The machine's opinion
is unaltered: 
$$ 1/2 \ket{0} \bra{0} + 1/2 \ket{1} \bra{1} \, . $$ 
Aha!  The machine's failed, as expected.  The qubit is now in a pure
state, not a mixture.  You explain this, and your arrangement with 
Isabelle, to Zweisteine, who listens intently, and asks you nonetheless
to continue.  

So, at $12.59$pm, you feed the qubit in again, and 
again read 
$$ 1/2 \ket{0} \bra{0} + 1/2 \ket{1} \bra{1} \, . $$ 
At $1.01$pm you try once more, and for the second time that day
are surprised by the printout: 
$$ c \ket{0} + d \ket{1} \, , $$
an opinion which the machine maintains as you desultorily resubmit the
qubit over the next half hour.
When Isabelle's radio message arrives at $1.30$, you find she measured in the 
basis $ c \ket{0} + d \ket{1} , \bar{d} \ket{0} - \bar{c} \ket{1}$,
and obtained the second state.  

This can't be fraud.  Isabelle and Zweisteine have never met, and anyway
she is entirely trustworthy.  You remind yourself that, for all his
eccentricities, and despite his scandalous neglect of the quant-ph
arxiv, Zweisteine is a dedicated scientist, and a good one.  He has
been exploring unfamiliar physics, ranging from quantum effects in
neurophysiology and consciousness to strong-field gravity, and
not without success.  In fact, some recent effects he's discovered
are said by experts to be inexplicable by conventional theory. 
And his lab has, come to think of it, lately 
taken delivery of some specially bioengineered neural circuits and
premium grade black holes. 

You begin to reconsider$\ldots$

\section{What could a pure state readout device describe?} 

Zweisteine's machine appears to be functioning as a genuine
quantum state readout machine for pure states.  When presented
with a state of an entangled subsystem, it appears to recognise 
that it is entangled.  However, it is apparently unaware of 
distant measurements that disentangle the state, until the
point when information about those measurements could have
reached it by light speed communication.  What principles
could it be following, consistent with quantum theory and 
relativity?  

To simplify the notation, consider distinguishable pointlike particles
located at fixed points $\ul{x}_1, \ul{x}_2 , \ldots , \ul{x}_N$ in
some inertial coordinate system $( \ul{x}, t)$, and that the
particles' spatial wave function spread is negligible throughout the
following discussion.  The particles have some internal degrees of
freedom, and their joint state is, we'll assume, entangled at $t=0$:
$$
\ket{\psi (0) } = \sum_{i_1 \ldots i_N} a_{i_1 \ldots i_N} \ket{i_1}_1
\ldots \ket{i_N}_N  \, . 
$$
Suppose also that the particles have no mutual interactions and
have been undisturbed, prior to $t=0$, for a time long compared
to their spatial separation, and remain so up to time $t_1 > 0$:
$$ \ket{ \psi (t) } = \ket{\psi (0) } {\rm~for~} -T < t < t_1 \, , 
$$
where $T \gg \max_{i,j} ( \| \ul{x}_i - \ul{x}_j \| )$. 

What is the state of particle $1$ at $t=0$?  The standard textbook answer is 
that it has no pure state, but is in an (improper) mixed state:
$$
\rho_1 (0) = \Tr_{2 , \ldots , N} ( \ket{\psi(0)} \bra{\psi(0)} ) \, . 
$$

We now want to consider how measurements affect the state. 
It will be assumed that measurement is an objectively definable 
process, and that a genuine state vector reduction takes place during
measurement.  That is, the quantum state of the measured system 
alters to one of the possible measurement outcomes; it does 
not enter into an entangled superposition
with the apparatus which includes all the possible results.  
Of course, this is not everyone's favoured approach to the
measurement problem.  But it is one of the standard options. 
The aim here is to explore the scope for hypothetical readout devices 
and nonlinear theories under the assumption that it is correct.  

Suppose now that a projective measurement is carried out on particle $2$
at time $t_1 > 0$, and it is found to be in state $\ket{j}_2$.  
Write $P^j = \ket{j} \bra{j}$ and 
$P^j_2 = I \otimes P_j \otimes I \otimes \ldots \otimes I$.  
The textbook version of the projection postulate, the state
of the full system is now, up to a normalisation factor,  
$$ \ket{ \psi (t_1)} = P^j_2 \ket{ \psi (0) } \, ,$$
and the state of particle $1$ is now, again up to normalisation,
$$
\rho_1 (t_1 ) = \Tr_{2 , \ldots , N} ( \ket{\psi(t_1 )} \bra{\psi( t_1
  )} ) \, . 
$$
Generally, $\rho_1 (t_1 )$ and $\rho_1 (0)$ will be different.
On this account, the state of particle $1$ has instantaneously changed 
as a result of a distant measurement on particle $2$.  

Of course, had we used a different reference frame, we would
have found the state of particle $1$ changing instantaneously
at a different point on its worldline.  
Hence, famously, we cannot consistently maintain both the
relativity principle and that the
state of particle $1$ --- as defined by these calculations ---
represents an objective physical fact about the particle.
In particular, if we go further and postulate a hypothetical
device that reads out the value of the state as we have defined
it, we need to assume the device functions with respect to some 
preferred reference frame, and it then allows instantaneous
signalling in that frame over arbitrary distances. 

The dilemma pointed out by EPR, of course, is that there is
a plausible-seeming reason to think that the 
physical state of particle $1$ really might be 
objectively defined by $\rho_1 (t_1 )$, not
$\rho_1 (t_0 )$ after the measurement (and so one might think 
any sensible hypothetical state readout device {\it should} 
output $\rho_1 (t_1)$ after the measurement ).  Namely, 
measurements on particle $1$ after time $t_1$ have outcome
probabilities in accordance with $\rho_1 (t_1 )$, not
$\rho_1 (t_0 )$, and so the state of the particle, which is supposed 
to be the best available physical description, should be
$\rho_1 (t_1 )$.  
But then the relativity principle suggests
the state of particle $1$ should have been $\rho_1 (t_1 )$ 
{\it before} time $t_1$.  This leads us to introduce a local
hidden variables hypothesis, and then Bell's theorem seems to refute
this whole line of thought.  

Could there, though, be a genuinely objective description
of each of the particles that is weaker --- in the sense that
it is not always sufficient to reproduce all the predictions of quantum
theory --- but {\it is} consistent with relativity?   
Yes: in fact, there is a natural candidate, defined as follows.  
(See Figure 1.) 
\begin{figure*}
\includegraphics[width=4in,keepaspectratio=true]{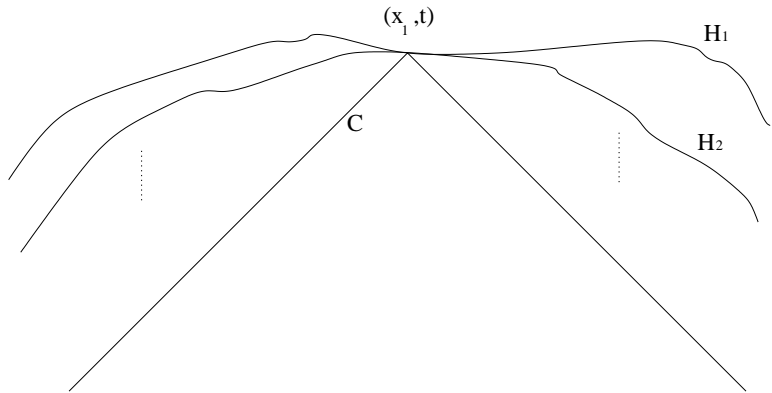}
\caption{Spacelike hypersurfaces tending to the past light cone}
\end{figure*}

Consider $C$, the surface of the past light cone of 
particle $1$ at time $t$.
Take a family $\{ H_n : n = 1, 2, \ldots \}$ of spacelike 
hypersurfaces which go through $( \ul{x_1 }, t )$ and which
asymptotically tend to $C$.  Let $\ket { \psi^n }$ be the 
state vector of the joint system on $H_n$, and define
$$\rho^n = \Tr_{2 , \ldots , N} ( \ket{\psi^n } \bra{\psi^n } )
\, . $$
Finally, define the {\it local state} of particle $1$ at time $t$ to be
$$ 
\rho^{loc}_1 = \lim_{ n \rightarrow \infty } \rho^n \, . 
$$
In words: the particle's local state is given by taking the joint wave
function of the complete system, defined by allowing for only
those projective measurements in the past light cone of the 
particle, and then tracing out the rest of the system. 

Clearly, $\rho^{loc}_1$ is Lorentz invariant.  It also has a natural
physical interpretation: $\rho^{loc}_1 (\ul{x_1}, t)$ is the best 
possible description of the state obtainable by an observer 
located at $( \ul{x_1}, t )$.  Such an observer can obtain $\rho^{loc}_1$ 
by knowing the initial state and having arranged for radio signals 
of all measurement outcomes to be sent to him as soon as the 
measurements take place: he will thus have details of all measurements
within the past light line cone of $( \ul{x_1 }, t )$.  

\section{Local state readout does not allow superluminal signalling} 

The above construction of $\rho^{loc}$ has another significant
implication.  Assuming that standard quantum theory is correct, that
we know the initial state of a system, and that we can identify all
measurement events on that system and obtain their results, we could
in principle construct a local state readout machine emulator for the
system --- that is, a device that will have the same operational
action as a local state readout device for local subsytems.  

To do this would require complete information about the system's
hamiltonian and ideal technology --- communication devices set up
everywhere that broadcast signals reporting the results of measurements
at light speed in all directions, and computers set up everywhere
that carry out arbitrarily fast calculations.  
Given these things, and the value of the initial state, we can program the
computers to take account of all measurement results as soon as 
the signal reporting them arrives, and use these together with 
knowledge of the hamiltonian evolution in the past light
cone to calculate the local state and print it out.  All of 
this can be done classically: the computers do not need to 
carry out any additional measurements on the system or disturb 
its quantum state in any way.  Hence they emulate the state
readout device, as required, by producing the state's value
while leaving it undisturbed.   

Obviously, these assumptions are unrealistic.  We do not know
the initial state of the universe, nor can we identify all
measurements in our past light cone, nor can we construct
the ideal technology required.   But none of these assumptions
contradicts standard quantum theory: each of them can consistently
be added to it without changing the underlying theory.  
Hence, since standard quantum theory does not allow superluminal
signalling, nor does quantum theory augmented by devices which
emulate local state readout machines.  And since there is no
operational distinction between a local state readout machine
emulator and a local state readout machine, quantum theory augmented
by genuine local readout readout machines does not allow
superluminal signalling either (happily for Zweisteine).

\section{Emulating nonlinear theories} 

Given the hypothesis of local state readout machines, we can go
further and devise experiments in which the hamiltonian acts on
the quantum state as usual defined, but is defined in terms
of fields which depend locally on the local quantum state.
To construct such experiments, we would simply need to 
connect the readout to another device which controls an applied
field.   For instance, given a system of separated qubits and
with some fixed basis,
we could arrange for the hamiltonian to include a term 
$\pi / 4 \bra{0} \rho_{\rm loc} \ket{0} \sigma_z$.  
More generally, we could implement
any locally varying 
nonlinear evolution laws of our choice {\it provided that the
nonlinearity arises through dependence on the local state}.   

Now, we have already seen
that a device operationally indistinguishable from a local state 
readout device could be constructed within standard quantum 
theory, given sufficient knowledge and computational power, 
and hence that such a device does not allow superluminal signalling.
It follows that superluminal signalling cannot be possible in any
experiment of this type.  But these experiments emulate a situation
in which nature (through presently unknown physics) uses locally
varying nonlinear evolutions that depend on the local state.
Hence no theory of this type can allow superluminal signalling either.  

\section{Conclusions} 

Could Zweisteine be right?  Might unknown physics
give a direct way of carrying out local state readout, or at
least some partial information about the local state, on 
general quantum systems?   Despite the lessons of Bell's 
theorem and experimental verifications of quantum nonlocality,
there is still some attraction in the idea that there is 
{\it something} objectively ``there'' in a localised part of 
an entangled quantum system.  If not the local state, what?   

Suppose, for instance, that, as has sometimes been speculated, that 
the gravitational field is actually fundamentally classical, while
matter is quantum.  The gravitational field then has to couple to 
some object defined by the quantum realm, and the local state seems 
a plausible candidate.  One might also wonder whether a theory of
consciousness, which (according to one line of thought) has to 
attach consciousness to some definite physical quantity, might
possibly use local quantum states.  

The problem, of course, in taking these thoughts beyond coffee table
speculation into specific detail is that infinitely many local state
dependent evolution laws could be written down.  One of the initial
hopes --- that requiring consistency with special relativity might
reduce the number of nonlinear theories to a few candidates --- has
not been fulfilled.  Perhaps it might be possible to identify a
restricted class of sensible ans\"atze for coupling the local state to
gravity, though.

These speculations aside, the fact remains that a theory which
implies nonlinear evolution of pure quantum states need not
allow superluminal signalling, or otherwise violate relativity.  
With this concern lifted, testing quantum linearity seems a more
respectable enterprise than it has lately been painted.

\section{Acknowledgements} 

This work was funded in part by a Royal Society University 
Research Fellowship, the European project EQUIP and the Cambridge-MIT
Institute.  I thank the Perimeter Institute for support while
the final version of the paper was completed.   
Zweisteine thanks Harvey Brown's character Keinstein \cite{brown}
for helpful conversations.

\end{document}